# AI and Accessibility: A Discussion of Ethical Considerations

Meredith Ringel Morris
Microsoft Research
merrie@microsoft.com

According to the World Health Organization, more than one billion people worldwide have disabilities. The field of disability studies defines disability through a social lens; people are disabled to the extent that society creates accessibility barriers. AI technologies offer the possibility of removing many accessibility barriers; for example, computer vision might help people who are blind better sense the visual world, speech recognition and translation technologies might offer real time captioning for people who are hard of hearing, and new robotic systems might augment the capabilities of people with limited mobility. Considering the needs of users with disabilities can help technologists identify high-impact challenges whose solutions can advance the state of AI for all users; however, ethical challenges such as inclusivity, bias, privacy, error, expectation setting, simulated data, and social acceptability must be considered.

**Inclusivity**
The inclusivity of AI systems refers to whether they are effective for diverse user populations. Issues regarding a lack of gender and racial diversity in training data are increasingly discussed; however, inclusivity issues with respect to disability are not yet a topic of discourse, though such issues are pervasive [4, 16]. For example, speech recognition systems, which have become popularized by virtual assistants, do not work well for people with speech differences such dysarthria, deaf accent, etc., since training data does not typically include samples from such populations [6]. Advances in computer vision have led several groups to propose using such algorithms to identify objects for people who are blind, but today's vision-to-language algorithms have been trained on data sets comprised of images taken by sighted users, limiting their efficacy when applied to images captured by blind users, which tend to have far lower quality [2]. These inclusivity issues threaten to lock people with disabilities out of interacting with the next generation of computing technologies. Proposed methodologies for increasing awareness of the provenance and limitations of data sets [3] are a starting point to increasing data set transparency. Efforts to directly source data from under-represented user groups, such as the VizWiz data set [5], which contains thousands of images and related questions captured by people who are blind, are a step in the right direction.

**Bias**
Just as AI technologies can amplify existing gender and racial biases in our society, they can also exacerbate disability-based discrimination. Recent research indicates that AI researchers can infer disability status from online data traces; for example, our research showed we could infer whether people are blind by analyzing their Twitter profiles and activity [10], and researchers have also demonstrated the ability to assess whether people have Parkinson's disease based on their mouse movements while on a search engine homepage [18]. The potential for peoples' disability status to be implicitly revealed through their computing actions means that algorithms could treat users differentially based on inferred disability status, such as the possibility for health insurers to deny coverage or employment advertisements to be targeted to avoid display to people with disabilities [7]. Developing ethical and legal frameworks regarding the application of AI to inferring disability status is a crucial issue.

**Privacy**

People with rare disabilities may experience greater privacy risks if they contribute data to AI systems or participate in research studies evaluating AI technologies. Past incidents of re-identification of individuals from anonymized data sets, such as the 2006 AOL search data leak, indicate the difficulty of truly anonymizing data. For some small disability populations, privacy-preserving techniques such as k-anonymity may not be effective, creating increased re-identification risk. These increased privacy risks to people with disabilities are amplified by the aforementioned *bias* issues that people with disabilities may face if their disability status is exposed, and creates an incentive for people with disabilities to withhold their data from research studies, an issue that can further amplify the *inclusivity* problem of AI systems. Hence, reflection on how current research practices may impact risks of deductive disclosure (e.g, [1]), as well as development of stronger technical and legal privacy frameworks are critical to creating accessible AI technologies.

**Error**

Many people with disabilities need to trust and rely on the output of an AI system without the ability to verify the output (e.g., a person who is blind relying on the output of a computer vision system). Our recent study found that people who were blind were over-trusting of an AI image captioning system, even when the output made little sense [9]; we also found that incorrect output from a computer vision system negatively impacted the understanding of blind users to an extent that could not be corrected even with human assistance [13]. Precision/recall tradeoffs for AI systems may need to be re-calibrated for vulnerable user populations – for example, the level of confidence a computer vision system must have to offer a label of an image for retrieval by sighted users of an image search engine may be quite different than the level of confidence the same system would need to offer a label of a scene to a blind pedestrian, where errors may have safety consequences. Error metrics output by AI systems must also be intelligible to end users. Translating mathematical confidence values to user-actionable information is an open challenge for AI that is particularly critical to users with disabilities.

**Expectation Setting**

The language used to describe AI technologies in the scientific literature, the popular press, and advertising materials often creates unreasonable expectations of such systems' current capabilities to lay users [8]. For example, an article announcing that machine translation systems have reached "human parity" for very specific tasks on very specific data sets may lead the general public to make incorrect assumptions about the current state of the art of such AI in the open world. Terminology, demos, articles, and marketing materials that lead the public to misunderstand the capabilities of AI systems are particularly ethically problematic in the case of sensitive user populations, such as people with disabilities, whose quality of life may be fundamentally changed by such advances. In the domain of healthcare, bodies like the FDA regulate the kinds of statements that can be made about the efficacies of various treatments, yet no such regulations exist around promises made regarding the capabilities of apps and algorithms that may impact users' health, well-being, and ability to perform activities of daily living.

**Simulated Data**

As the need to create inclusive data sets is increasingly recognized, some technologists have used simulation to generate synthetic data; for instance, one approach has been to digitally modify or generate data to create variation not originally present in a data set [15]. Simulating disability is generally not recommended [14], since studies have found the data generated by users simulating disabilities (e.g., a sighted person wearing a blindfold) is not the same as that from people who are truly disabled (i.e., a person who has been blind for years will be much more skilled than a person who has just put on a blindfold). Further, disability simulation has been found to create negative impressions of the capabilities of people with disabilities [11]. However, lack of data is a tension that makes simulation appealing to many technologists. For example, AAC (augmentative and alternative communication) systems that are used by many people with severe speech and motor disabilities are often extremely slow to operate; intelligent language prediction can greatly increase communication bandwidth for AAC users. However, status quo prediction models are often trained on publicly available corpora such as news articles, whose linguistic structure may not represent the typical speech patterns of the target end users. To address this problem, researchers created a simulated AAC speech corpus by asking workers on Amazon's Mechanical Turk service to imagine what they might say if they were disabled [17]; the resulting corpus is skewed toward stereotyped phrases such as "Can I have some water please?" or "Who will drive me to the doctor's office tomorrow?", despite the fact that actual AAC users wish to talk about interests and hobbies beyond their healthcare needs [6]. Creating guidelines and best practices around the use of simulation for developing and testing AI systems for people with disabilities is important for ensuring that systems are not trained on non-representative data. Solutions addressing the aforementioned *privacy* challenges, which may prevent some people with disabilities from contributing data to AI efforts, may be particularly important for enabling inclusive data set creation that obviates the need for data simulation.

**Social Acceptability**
The deployment of AI systems in the open world creates important questions surrounding the social acceptability of new technologies. Of particular concern are questions regarding technologies' impact on indirect stakeholders, including issues of privacy and fairness. As an example of how privacy concerns influence social acceptability, consider the case of wearable technologies that may be capturing visual and audio data for algorithmic processing, such as Google Glass, which faced a public backlash due to privacy concerns. However, recent research indicates that people are more tolerant of such systems if the technology is being used by a person with disabilities [12]. Technologists, ethicists, and lawmakers must grapple with whether and how rules regarding the use of AI systems might change depending on an end user's disability status, and, if such differential uses of technology are deemed acceptable, how such distinctions could be facilitated in practice.

**Conclusion**
AI technologies offer great promise for people with disabilities by removing access barriers and enhancing users' capabilities. However, ethical considerations must carefully guide the development, deployment, and discussion around such technologies. These considerations regarding inclusivity, bias, privacy, error, expectation setting, simulated data, and social acceptability apply to all users, but are particularly nuanced and salient when considering the large potential benefits and large potential risks of AI systems for people with disabilities. While legal regulation may address some of these considerations in the future, it is unlikely to keep pace with the changing technological landscape. Educating our next generation of innovators is of paramount importance; emerging ethics curricula for computer science students should include content such as sociological concepts from the field of disability studies and inclusive design approaches from HCI. It is also vital to ensure that people with disabilities have voice in technological innovation by making our educational institutions and workplaces more inclusive so as to increase the representation of people with disabilities among computer science professionals. As technologists, it is our responsibility to pro-actively address these issues in order to ensure that people with disabilities are not left behind by the AI revolution.

**Author Bio**
Dr. Meredith Ringel Morris is a Sr. Principal Researcher at Microsoft Research, where she leads the Ability team in conducting research that combines advances in HCI and AI to address the technology needs of people with disabilities. She is also an affiliate Professor at the University of Washington.